\documentstyle[epsfig,aps]{revtex}
\def\nn{\nonumber}
\def\beq{\begin{equation}}
\def\eeq{\end{equation}}
\def\bea{\begin{eqnarray}}  \def\eea{\end{eqnarray}}
\def\lsim{\raise0.3ex\hbox{$<$\kern-0.75em\raise-1.1ex\hbox{$\sim$}}}
\def\gsim{\raise0.3ex\hbox{$>$\kern-0.75em\raise-1.1ex\hbox{$\sim$}}}
\def\1{{\rm 1\mskip-4.5mu l} }
\newcommand{\noi}{\noindent}
\begin{document}
\flushbottom
\def\thepage{\roman{page}}
\title{\vspace*{1.5in} CHARGED MULTIPLICITIES AND $J/\psi$
SUPPRESSION AT SPS AND RHIC ENERGIES}
\author{A. Capella and \underline {D. Sousa}}

\address{Laboratoire de Physique Th\'eorique, UMR 8627 CNRS,\\
Universit\'e de Paris XI, B\^atiment 210, 91405 Orsay Cedex, France}

\maketitle
\begin{abstract}
Charged multiplicities in nucleus--nucleus collisions are calculated
in the Dual Parton Model taking into account shadowing corrections.
Its dependence on the number of collisions and participants is analyzed
and found in agreement with experiment at SPS and RHIC energies.
Using these results, we compute the $J/\psi$ suppression at SPS
as a function of the transverse energy and of the energy of the zero
degree calorimeter.
Predictions for 
RHIC are presented.
\end{abstract}

\section*{Charged Multplicities in the Dual Parton Model}

In the Dual Parton Model (DPM) the charged multiplicity per unit rapidity
in a symmetric collision is given by \cite{1r}

\begin{eqnarray}
\label{1e}
{dN_{AA}^{ch} \over dy}(y,b) & = & n_A (b) \left [ N_{\mu}^{qq^{P}-q_v^T}(y) +
N_{\mu}^{q_v^P-qq^T}(y) + (2k - 2) N_{\mu}^{q_s-\bar{q}_s} \right ] + \nn \\
&&\Big ( n(b) - n_A(b)\Big ) \Big ( 2 k \ N_{\mu}^{q_s-\bar{q}_s}(y)
\Big ) \quad .
\end{eqnarray}

\noi  Here $P$ and $T$ stand for the projectile and target nuclei,
$n(b)$
is the average number of binary collisions and $n_A (b)$
the average number of participants of nucleus $A$. These quantities
can be computed in a Glauber model. $k$ is
the average number of inelastic collisions in $pp$ and
$\mu (b) = kn(b)/n_A(b)$ is the total average number of collisions suffered
by each nucleon. The first term in (\ref{1e}) is the plateau
height in a $pp$
collision, resulting from the superposition of $2k$ strings,
multiplied by $n_A$. Since in DPM
there are two strings per inelastic collision, the second term,
consisting of strings stretched
between sea quarks and antiquarks, makes up a total number of strings
equal to $2kn$. \par

The charged multiplicity produced by a single string is obtained
by a convolution of
momentum distribution function and fragmentation functions
(eqs. (3.1) to (3.4) of \cite{2r}).

Shadowing corrections in Gribov theory are universal \cite{3r}, i.e. 
they apply both
to soft and hard processes. 
The reduction
of the multiplicity resulting from shadowing corrections has been computed in
\cite{3r}. These
corrections are negligeable at SPS energies
but at RHIC energies they reduce the multiplicity by 40 to 50\%.

We present the results \cite{1r} obtained at two
different energies~: $\sqrt{s} = 17.3$ and 130 GeV. The corresponding
non-diffractive cross-sections are $\sigma_{ND} = 26$ and 33 mb,
respectively. We take $k = 1.4$ and 2.0 corresponding to $dN_{pp}^{ND}/dy = 1.56$
and 2.72. The result in absence of shadowing
at $\sqrt{s} = 17.3$ is shown in Fig~1 (a).
We obtain a mild increase of the multiplicity per participant
consistent with the results of the WA98 Collaboration \cite{3br}.
This increase gets stronger with increasing energies. As we pointed out before,
shadowing corrections are negligeable at SPS energies but their effect
is large at RHIC. Unfortunately, shadowing corrections have a rather large
uncertainty at RHIC energies. Two alternative calculations \cite{1r} of
shadowing lead to the results at $\sqrt{s} = 130$ GeV
shown by the solid lines in Fig~1 (b).
Clearly, with the larger values of the
shadowing corrections we obtain a 
quantitative agreement with the PHENIX data \cite{3cr}.

\begin{figure}[t]
\begin{center}
\begin{minipage}{40mm}
\mbox{\epsfig{file=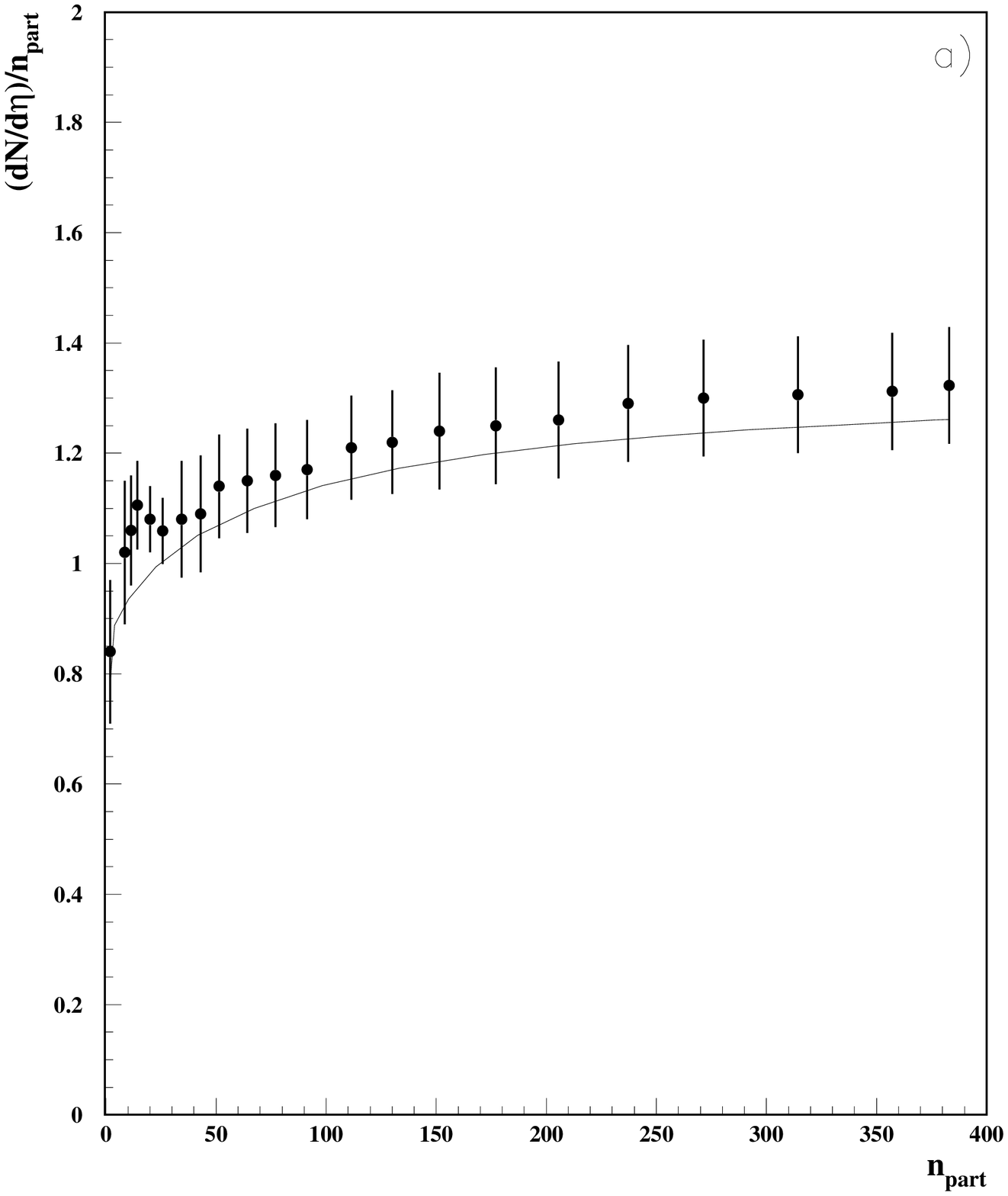,bbllx=0,bblly=30,bburx=480,bbury=620,height=2.8in}}
\end{minipage}
\hspace{\fill}
\begin{minipage}{40mm}
\mbox{\epsfig{file=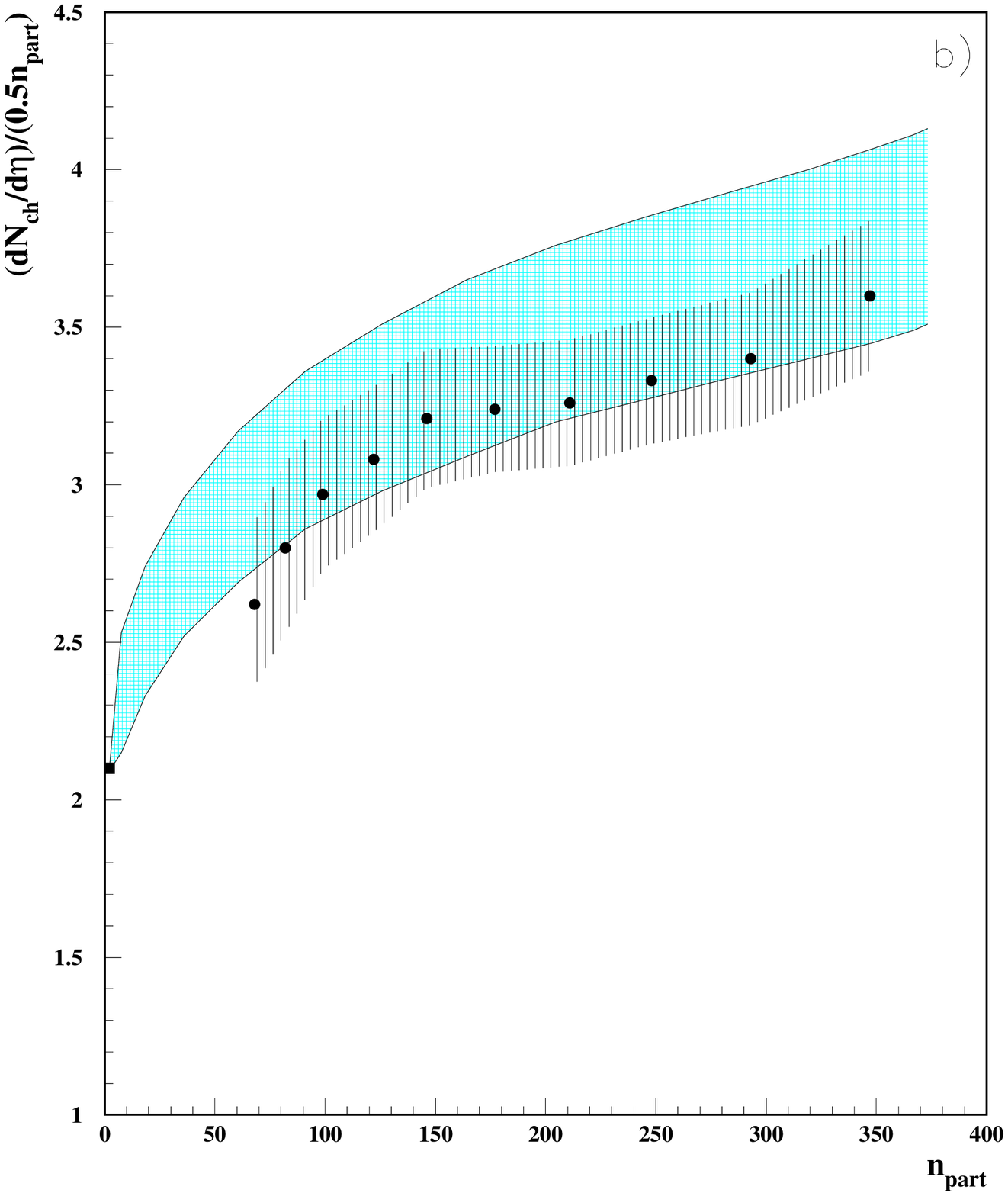,bbllx=280,bblly=30,bburx=840,bbury=620,height=2.8in}}
\end{minipage}
\end{center}
\caption{
a) The values of $dN^{ch}/d \eta /n_{part}$
versus $n_{part}$ for
$PbPb$ collisions at $\sqrt{s}
= 17.3$~GeV in the range $- 0.5 < \eta_{cm} < 0.5$ computed from eq.
(\ref{1e}), compared with the
WA98 data \protect\cite{3br}.
b) The values of $dN^{ch}/d \eta /(0.5 n_{part})$ for
$Au$-$Au$ collisions at
$\sqrt{s} = 130$~GeV in the range $- 0.35 < \eta_{cm} < 0.35$ computed
from eq. (\ref{1e}) taking into account shadowing corrections (see main
text). They are compared to the data \protect\cite{3cr}.}
\end{figure}

Note that our calculations refer to $dN/dy$ while
the first RHIC measurements refer to $dN/d\eta$.
The latter is, of course, smaller at mid rapidities. This
difference is negligibly small as SPS where the laboratory
pseudo-rapidity variable is used. However, at $\sqrt{s} = 130$ GeV
where $\eta_{cm}$ is used instead, their ratio can be as large as 1.3 \cite{3dr}.

The predictions at $\sqrt{s} = 200$ GeV were also given in \cite{1r}.
The predicted increase between $130$ and $200$ GeV is $13$~\% in quantitative
agreement with the measurement by the Brahms Collaboration \cite{3er}.

\section*{$J/\psi$ suppression vs $E_{T}$ and $E_{ZDC}$ at SPS}

In the comovers approach the $J/\psi$ survival
probability is the product of two factors $S_{abs}(b,s)\cdot S_{co}(b,s)$. The
first factor represents the suppression due to nuclear absorption of the
$c\bar{c}$ pair. Its expression, given by the probabilistic Glauber model, is
well known. It contains a parameter, the absorptive cross-section
$\sigma_{abs}$. The second factor $S_{co}(b,s)$ represents the suppression
resulting from the interaction with comovers.
They
are given by

\beq
\label{6e}
S^{abs}(b, s) = {[1 - \exp (- AT_A(s) \ \sigma_{abs})] [1 - \exp (- B
\ T_B (b - s) \ \sigma_{abs})]
\over \sigma_{abs}^2 \ AB \ T_A(s) \ T_B(b - s)} \eeq

\beq
\label{7e}
S^{co} (b, s) = \exp \left [ - \sigma_{co} {3 \over 2}
N_{y_{DT}}^{co}(b, s) \ell n \left ( {{3
\over 2} N_{y_{DT}}^{co}(b, s) \over N_f} \right ) \right ]  \eeq

\noi In (\ref{7e}), $N_{y_{DT}}^{co}(b, s)$ is the density of charged
comovers (positives and
negatives) in the rapidity region of the dimuon trigger and $N_f =
1.15$~fm$^{-2}$ \cite{5r} \cite{6r}
is the corresponding density in $pp$. The factor 3/2 in (\ref{7e})
takes care of the neutrals.
In the numerical calculations we use
$\sigma_{abs} = 4.5$~mb
and
$\sigma_{co} = 1$~mb \cite{7r} \cite{8r}. We compute the density of
comovers in the framework of the DPM as we have explained before.
\par

This approach allows to compute the impact parameter of the $J/\psi$
event sample. Experimental results of this quantitie are plotted
as a function of observable quantities such as $E_{T}$ or $E_{ZDC}$.
Using the proportionality between $E_{T}$ and multiplicity, we
have 

\beq
\label{7be}
E_{T}(b) = {1 \over 2} q N^{co}_{y_{cal}}(b) \quad .
\eeq

\noi The multiplicity of comovers $N_{y_{cal}}^{co}(b)$ is determined
using Eq. (\ref{1e}) in the rapidity region of the NA50 calorimeter
($1.1 < y_{lab} <2.3$). The factor $1/2$ is introduced because 
$N_{y_{cal}}^{co}(b)$ is the charged multiplicity whereas $E_{T}$
refers to neutrals. Thus the coefficient $q$ is close to the average 
energy per participant and its value can be determined from the position
of the "knee" of the $E_{T}$ distribution of the $MB$ events measured by
the NA50 Collaboration. We obtain $q = 0.62$ GeV \cite{8br}.

The energy of the zero degree calorimeter is defined as

\beq
\label{7ce}
E_{ZDC}(b) = [A-n_{A}(b)]E_{in} + \alpha n_{A}(b) E_{in} \quad .
\eeq

\noi Here $A-n_{A}(b)$ is the number of spectator nucleons of $A$ and
$E_{in} = 158$ GeV is the beam energy. The last term represents the 
small fraction of wounded nucleons and/or fast secondaries that hit
the ZD Calorimeter. The value of $\alpha$ can be precisely determined
from the position of the $MB$ event sample measured by NA50. 
We obtain \cite{8cr}
$\alpha = 0.076$. Eqs. (\ref{7be}) and (\ref{7ce}) also lead to a
correlation between (average values of) $E_{T}$ and $E_{ZDC}$. This
correlation is close to a straight line \cite{8cr} and 
gives a good description of the
experimental one.

To explain the experimental data beyond the knee of the $E_{T}$
distribution we introduce two effects:

1. {\bf Comovers fluctuations} \cite{7r}: We introduce the fluctuation in
the density of comovers 
by replacing $N_{y_{DT}}^{co}$ in Eq. (\ref{7e}) by
$N_{yDT}^{co}(b,s)F(b)$ where $F(b)=E_{T}/E_{T}(b)$.
Here $E_{T}$ is the measured value of the trasverse energy and $E_{T}(b)$
is its average value given by Eq. (\ref{7be}) - which does not contain 
the fluctuations.

2. {\bf $E_{T}$ loss} \cite{8br}: In the $J/\psi$ event sample, $E_{T} \sim 3$
GeV is taken by the $J/\psi$ trigger and, thus, the transverse
energy deposited in the calorimeter by the other hadron species will be
slightly smaller than the corresponding one in the $MB$ event sample.

The results of our model for the ratio $J/\psi$ over $DY$, versus
$E_T$, in $PbPb$ collisions at $\sqrt{s} = 158$ GeV, are shown in
Fig.~2a and compared with NA50 data \cite{8dr,8gr} - both for the
true $J/\psi$ over $DY$ ratio, and for the one obtained with the $MB$
analysis. The results of the model for the true ratio are given by the
dotted line (without fluctuations) and the dashed line (with fluctuations).
In both cases the $E_{T}$ loss mechanism is not taken into account. 
However, our results for the true ratio $J/\psi$ over $DY$ do not
change, since the effect due to
the $E_T$ loss cancels in this ratio.
We see that our results are in good agreement with the NA50
which
do not extend beyond the knee.
The other data in
Fig.~2a are obtained with the $MB$ analysis, and have to be compared with 
the dashed-dotted and solid curves (obtained taking into account the
$E_{T}$ loss). In this case the agreement with the NA50 data is
substantially improved.

In Fig.~2b the results for the ratio $J/\psi$ over $DY$ versus
$E_{ZDC}$ are shown. These curves are obtained from the
corresponding ones versus $E_{T}$ applying the $E_{T}-E_{ZDC}$
correlation \cite{8cr}. 
Comparing the data with the model predictions, we see that
a better description of the central data (small $E_{ZDC}$) is obtained
when the $E_T$ fluctuations are taken into account. This was to be
expected since the fluctuations in $E_T$ and $E_{ZDC}$ are related to
each other via the $E_T - E_{ZDC}$ correlation. It is important to note
that the effect of the $E_{T}$ loss is not present in this case.
Indeed $E_{ZDC}$ measures the energy of spectators and it is not 
affected by the dimuon trigger. No disagreement between the data
and the model predictions is observed in this case for very central events.

\begin{figure}[t]
\begin{center}
\begin{minipage}{40mm}
\mbox{\epsfig{file=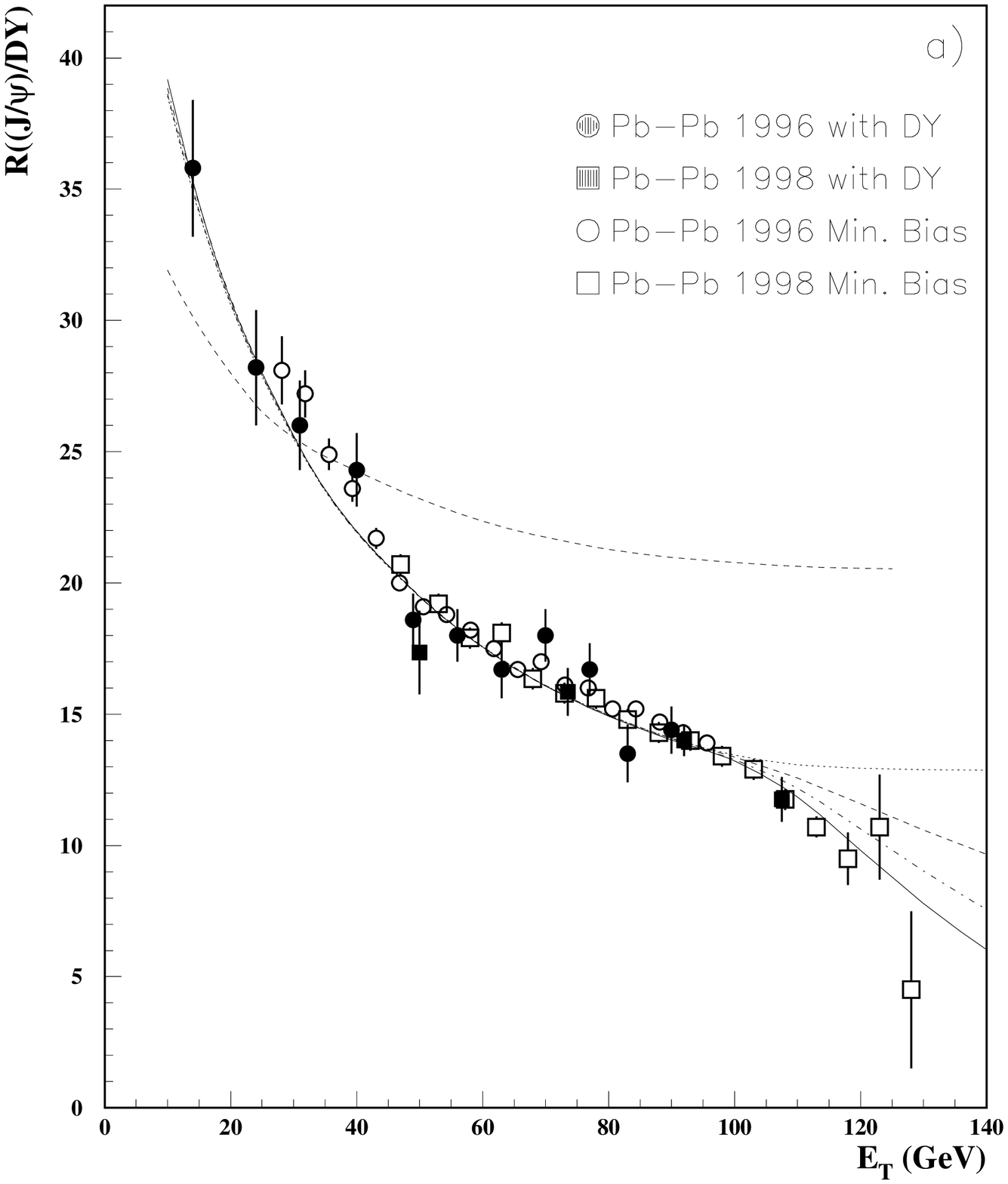,bbllx=0,bblly=30,bburx=480,bbury=620,height=2.8in}}
\end{minipage}
\hspace{\fill}
\begin{minipage}{40mm}
\mbox{\epsfig{file=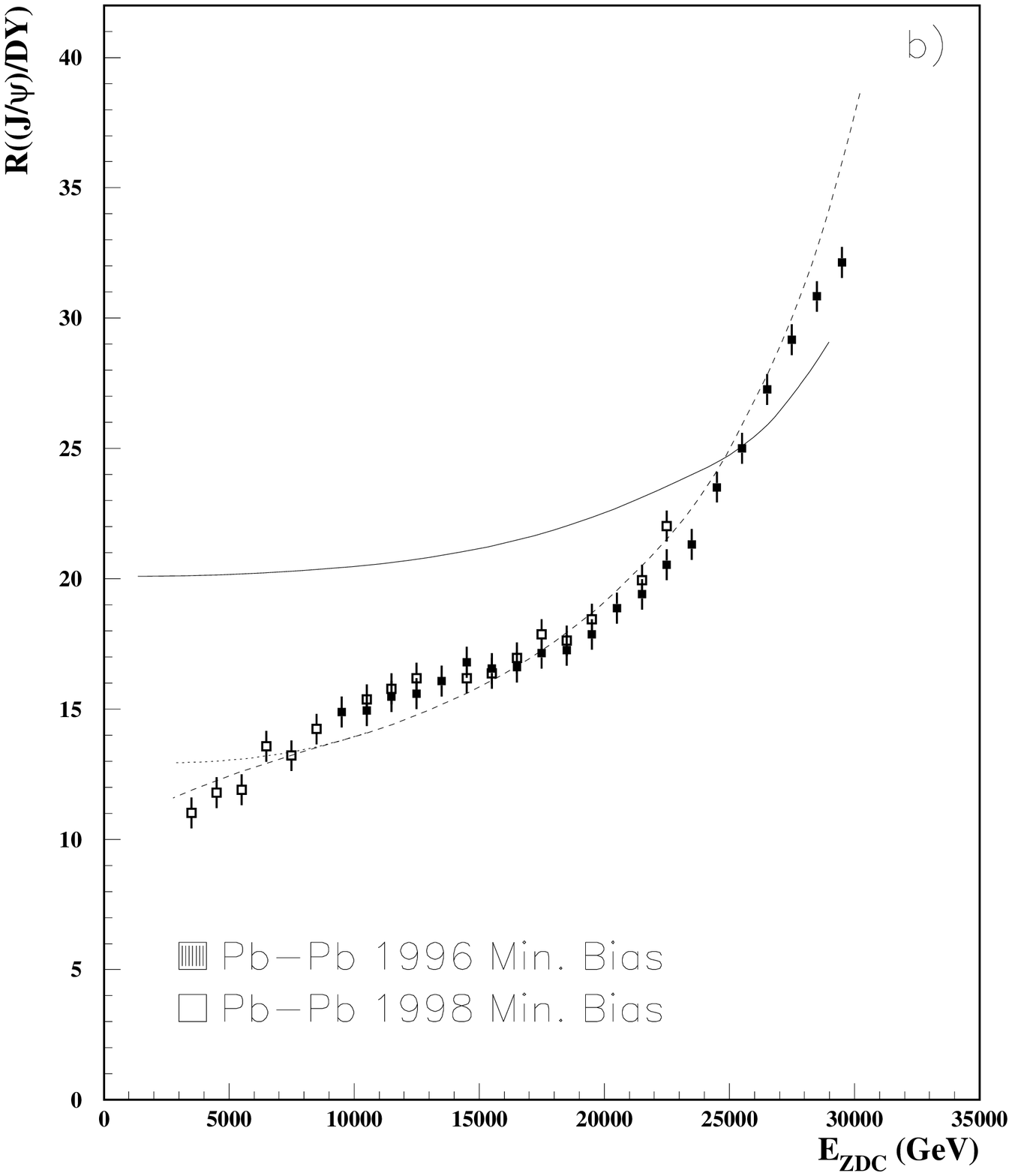,bbllx=280,bblly=30,bburx=840,bbury=620,height=2.8in}}
\end{minipage}
\end{center}
\caption{
a) Ratio $J/\psi$ over $DY$ versus $E_{T}$ in
$PbPb$ collisions at $158$ GeV \protect\cite{8br}. 
The data are from ref. \protect\cite{8dr}.
The data labeled with $DY$ are for the true $J/\psi$ over $DY$
and they should be compared with the
dotted and dashed lines
obtained, respectively, without and with $E_{T}$ fluctuations.
The data labeled Min. Bias should be compared with the dashed-dotted and
solid lines, obtained with the $E_{T}$ loss.
b) Ratio $J/\psi$ over $DY$ versus $E_{ZDC}$
in $PbPb$ collisions at 158 GeV per nucleon \protect\cite{8cr}.
The dashed (dotted)
curve is obtained from the dashed (dotted) one versus $E_T$
applying the $E_T - E_{ZDC}$ correlation and keeping the normalization
unchanged.
The data are from
\protect\cite{8dr} \protect\cite{8gr}. 
The NA50 nuclear absorption curve is also
shown.}
\end{figure}

\section*{$J/\psi$ suppression at RHIC}

In order to compute
the $J/\psi$ suppression at RHIC in the comovers approach we need to
know not only the effects of nuclear absorption and comovers
interaction, but also the effect of shadowing -- which becomes
important at RHIC energies. 
Therefore, our predictions are only valid so far as
the shadowing corrections cancel in the $J/\psi$ over $DY$ ratio --
which is not necessarily the case \cite{9r}. \par

For the comovers survival probability, Eq. (\ref{7e}), we keep the
value $\sigma_{co} = 1$~mb as discussed above. 
The hadronic multiplicity at RHIC energies
have been successfully evaluated in DPM \cite{1r}. Therefore, the
prediction at RHIC for this survival probability is rather safe. The
situation is quite different in what concerns the survival probability
due to nuclear absorption. It is widely 
recognized \cite{9r,10r,11r,12r,13r}
that, at
high energy, when the coherence length becomes larger than the nuclear
size, the probabilistic expression (\ref{6e}) is no longeer valid. It
has been shown in \cite{13r} that, at asymptotic energies, Eq.
(\ref{6e}) is replaced by

\beq \label{8e} S_{abs}(b, s) = \exp \left ( - {1 \over 2}
\ \widetilde{\sigma} \ A \ T_A(s) \right ) \exp \left ( - {1 \over 2}
\ \widetilde{\sigma} \ B \ T_B(b-s)\right ) \eeq

\noi where $\widetilde{\sigma}$ is the total $c\bar{c}$-$N$
cross-section. If $\widetilde{\sigma} \approx \sigma_{abs}$, the
asymptotic result is not very different from the one obtained with the
low energy formula -- since Eqs. (\ref{6e}) and (\ref{8e}) differ only in
the second correction term. This situation is expected if the
$c\bar{c}$ pair is produced in a colorless state interacting as a
dipole. However, the possibility has been advocated \cite{11r}, that
the $c\bar{c}$ pair is produced in a color state accompanied by light
quarks -- in order to make the system colorless. In this case the system
interacts with a comparatively large cross-section $\widetilde{\sigma}
\sim 15 \div 20$~mb \cite{10r,11r,14r}. With such a cross-section it
is possible to
explain the large suppression of the $J/\psi$ observed at $x_F \sim 1$,
without initial state energy loss of the projectile partons in the
nucleus. \par

\begin{figure}[t]
\begin{center}
\mbox{\epsfig{file=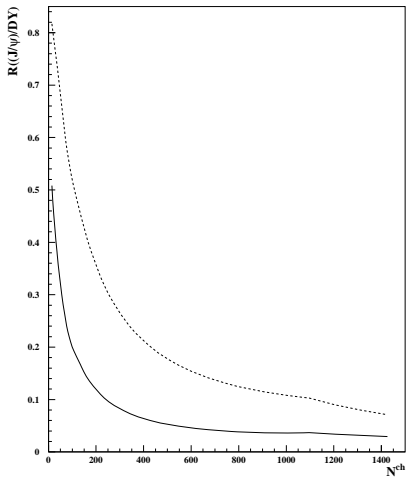,height=2.8in}}
\end{center}
\caption{Ratio $J/\psi$ over DY versus charged multiplicity in Pb-Pb
collisions at $\sqrt{s} =$ 200 AGeV in the range
$- 0.5 < y^* < 0.5$. The dotted line is obtained 
with the low energy nuclear
absorption Eq. (\ref{6e}) and $\sigma_{abs} = 4.5$ mb.
The full line is obtained with the 
asymptotic expression Eq. (\ref{8e})
and $\widetilde{\sigma} = 15$~mb.}
\end{figure}

The results for the ratio $J/\psi$ over $DY$ at mid-rapidities are
presented in Fig.~3. The dashed line is obtained with Eq. (\ref{6e})
and $\sigma_{abs} = 4.5$~mb. The solid line is obtained with Eq.
(\ref{8e}) and $\widetilde{\sigma} = 15$~mb. As we see, the difference
between the two predictions is very large. \par

The proposed way of measuring the $J/\psi$ suppression at RHIC is via
the ratio $J/\psi$ over $\Upsilon$. For the latter, the situation is
even more complicated. Due to its larger mass, the correlation
length is smaller and Eq. (\ref{8e}) is not valid even at
mid-rapidities. 
In this case a finite energy formula \cite{10r,12r,13r}
has to be used -- which interpolates between the low energy, Eq.
(\ref{6e}), and the asymptotic (Eq. (\ref{8e})) limits. 
Actually, even for $J/\psi$, Eq. (\ref{8e}) is not exact at
mid-rapidities. We have estimated that using the finite energy
formula with $\widetilde{\sigma} = 15$ mb the ratio $J/\psi$ over $DY$
for central events is 40~\% higher than the solid line in Fig.~3.
Therefore an
accurate measurement in $pA$ interactions will be necessary in order
to clarify the theoretical situation. \par

We thank N. Armesto, E. G. Ferreiro, A. Kaidalov and C. A. Salgado
for discussions. D. S. thanks Fundaci\'on Barrie de la Maza for
financial support.


\begin{references}
\bibitem{1r}A. Capella and D. Sousa, Phys. Lett. 
{\bf B511},185 (2001).
\bibitem{2r}A. Capella, U. Sukhatme, C-I. Tan, J. Tran Thanh Van,
Phys. Rep. {\bf 236},225 (1994). 
\bibitem{3r}A. Capella, A. Kaidalov, J. Tran Thanh Van, Heavy Ion
Physics {\bf 9} (1999). 
\bibitem{3br}WA98 collaboration, M. M. Aggarwal et al, nucl-ex/0008004. 
\bibitem{3cr}PHENIX collaboration, K. Adkox et al, nucl-exp/0012008.
\bibitem{3dr}P. Aurenche, F. Boop, A. Capella, J. Kwiecinski,
M. Marie, J. Ranft, J. Tran Thanh Van, Phys. Rev. {\bf D45}, 92 (1992).
\bibitem{3er}Brahms Collaboration, M. Baker in Proceedings International
Workshop on the Physics of the Quark-Gluon Plasma, Palaiseau, France,
4-7 Sepember 2001, to be published~; J. J. Gaardhoje, ibid.
\bibitem{5r}D. Kharzeev, C. Louren\c co, M. Nardi and H. Satz, Z.
Phys. {\bf C74}, 307 (1997).
\bibitem{6r}N. Armesto and A. Capella, Phys. Lett. {\bf B430},
23 (1998). N. Armesto, A.
Capella and E. G. Ferreiro, Phys. Rev. {\bf C59}, 359 (1999).
\bibitem{7r}A. Capella, E. G. Ferreiro and A. Kaidalov, Phys.
Rev. Lett. {\bf 85}, 2080 (2000).
\bibitem{8r}N. Armesto, A. Capella, E. G. Ferreiro, A. Kaidalov
and D. Sousa, nucl-th/0104004, Proceedings QM
2001, presented by A. Capella, ibid, and Proceedings XXXVI Rencontres de
Moriond, Les Arcs, France 2001, presented by D. Sousa.
\bibitem{8br}A. Capella, A. Kaidalov and D. Sousa, preprint LPT 01-42,
nucl-th/0105021.
\bibitem{8cr}A. Capella and D. Sousa, preprint LPT 01-91, 
nucl-th/0110072.
\bibitem{8dr}NA50 Collaboration, M. C. Abreu et al., Phys. Lett. 
{\bf B477}, 28 (2000).
\bibitem{8gr}NA50 collaboration, A. Romana in Proceedings International
Workshop on the Physics of the Quark-Gluon Plasma, Palaiseau, France,
4-7 Sepember 2001, to be published~; O. Drapier, ibid.
\bibitem{9r}B. Kopeliovich, A. Tarasov and J. H\"ufner, hep-ph/0104256.
\bibitem{10r}C. A. Salgado, hep-ph/01045231.
\bibitem{11r}K. Boreskov, A. Capella, A. Kaidalov and J. Tran Thanh
Van, Phys. Rev. {\bf D47}, 919 (1993).
\bibitem{12r}M. A. Braun and A. Capella, Nucl. Phys. {\bf B412}, 260 (1994).
\bibitem{13r}N. Armesto, M. A. Braun, A. Capella, C. Pajares and C.
A. Salgado, Nucl. Phys. {\bf B509}, 357 (1998).
\bibitem{14r}F. Arleo, P. B. Gossiaux, T. Gousset and J. Aichelin,
Phys. Rev. {\bf B61}, 054906 (2000).
\end{references}
\end{document}